# Eight-input optical programmable logic array enabled by parallel spectrum modulation

Wenkai Zhang[1][†], Bo Wu[1][†], Wentao Gu[1], Junwei Cheng[1], Hailong Zhou[1,*], Jianji Dong[1,*], Dongmei Huang[2,3], P. K. A. Wai[4] and Xinliang Zhang[1]

[1]Wuhan National Laboratory for Optoelectronics, School of Optical and Electronic Information, Huazhong University of Science and Technology, Wuhan 430074, China

[2]The Hong Kong Polytechnic University Shenzhen Research Institute, Shenzhen 518057, China

[3]Photonics Research Institute, Department of Electrical Engineering, The Hong Kong Polytechnic University, Hong Kong, 999077, China

[4]Department of Physics, Hong Kong Baptist University, Kowloon Tong, Hong Kong, 999077, China

[*]Corresponding author: hailongzhou@hust.edu.cn; jjdong@hust.edu.cn

[†]These authors contributed equally to this work

**Abstract:** Despite over 40 years' development of optical logic computing, the studies have been still struggling to support more than four operands, since the high parallelism of light has not been fully leveraged blocked by the optical nonlinearity and redundant input modulation in existing methods. Here, we propose a scalable multi-input optical programmable logic array (PLA) with minimal logical input, enabled by parallel spectrum modulation. By making full use of the wavelength resource, an eight-input PLA is experimentally demonstrated, and there are $2^{256}$ possible combinations of generated logic gates. Various complex logic fuctions, such as 8-256 decoder, 4-bit comparator, adder and multiplier are experimentally demonstrated via leveraging the PLA. The scale of PLA can be further extended by fully using the dimensions of wavelength and space. As an example, a nine-input PLA is implemented to realize the two-dimensional optical cellular automaton for the first time and perform Conway's Game of Life to simulate the evolutionary process of cells. Our work significantly alleviates the challenge of extensibility of optical logic devices, opening up new avenues for future large-scale, high-speed and energy-efficient optical digital computing.

## Introduction

Nowadays, the demand for computing is increasing rapidly, while the development of electronic computing faces great challenges of power consumption and latency, mismatching with the requirement of high-density computing. Compared with electronic computing, optical computing shows the potential advantages of high speed, low power consumption and high parallelism[1-3]. Since Boolean logic is the cornerstone of modern digital computing system, lots of attempts in optical implementation of logic operations have been made for more than 40 years[4-13]. Nonetheless, due to the nonlinear characteristic of logic computing, most optical logic schemes were confined to enable not more than four operands, such as two-input logic gate[14-16], 2-4 decoder[17-19], one or two-bit adder[20-22], comparator[23-25] and multiplier[26]. At this stage, it still faces significant challenges in supporting more operands, since the nonlinear requirement in optical logic severely hampers the optimal utilization of optical parallelism.

Among these schemes, the photonic programmable logic array (PLA) is one of the most representative logic computing works in light of its reconfigurability and universality. The existing PLA can be implemented by all-optical (AO) methods and electronic-optical (EO) methods. The AO methods harness nonlinear effects like four-wave mixing in highly nonlinear fiber[27,28] or semiconductor optical amplifier[29], to execute logic AND operations between input signals and generate the logic minterms. The limited parallelism and low efficiency of all-optical nonlinear effects hinder the scalability of all-optical PLA, and the latest work can only support three-input operands with a maximum of eight wavelengths[30]. The EO methods tend to generate logic minterms by controlling the on/off states of serial microring resonators based on the electro-optical nonlinear effect, which can be seen as optical logic AND operations between input electronic signals[31]. Despite the avoidance of all-optical nonlinearity, the output high level has multiple values and the number of required microring modulators is proportional to the square of the number of input operands. Limited by massy redundant input modulators, the number of input operands and the corresponding wavelengths in existing reports has not exceeded four[32,33]. The limitation of input operand of both methods severely impedes the further development and practical application of PLA in optical digital computing.

In this paper, we experimentally demonstrate an eight-input PLA with minimal logical input, enabled by parallel spectrum modulation. By configuring spectral modulators (SMs) with specific square-wave transmission spectrums, eight cascaded SMs can generate 256 logic minterms for eight input operands, corresponding to 256 different wavelength channels. Afterwards, a waveshaper (WS) is applied to select required wavelength channels thus accomplishing targeted logic functions. The possible combinations of 256 logic minterms amount to $2^{256}$, indicating our approach can serve as a general logic device for designing a wide variety of optical logic functions. Based on the proposed PLA, we implement an 8-256 decoder whose extinction ratio between all wavelength channels reaches more than 17dB. Other common logic functions like 4-bit comparator, adder and multiplier are also realized. Furthermore, we combine the dimensions of wavelength and space to achieve a nine-input PLA. The two-dimensional cellular automaton (CA) is realized on the optical platform for the first time, and the evolutionary process of each cell in Conway's Game of Life is accurately simulated with the PLA. The proposed PLA paves the way for large-scale, high-speed and energy-efficient optical digital computing.

## Results

### Principle

For simplicity, Fig. 1 depicts a three-input (Signal A, B, C) PLA consisting of three cascading SMs. To realize an SM, a WSS is first employed to partition the incoming spectrum into the positive ('+') and negative ('-') channels. And a 2×1 optical switch (OS) is then utilized to select the positive or negative channel according to input logical signal. This enables the SM to produce two complementary square-wave spectrums in the states of 0 and 1, respectively. Considering three input operands correspond to 8 logic minterms, the PLA necessitates 8 distinct wavelengths from the light source. After $SM_1$, Signal A is loaded on light at $\lambda_1$-$\lambda_4$ and Signal $\overline{A}$ is loaded on light at $\lambda_5$-$\lambda_8$. The period of the subsequent transmission spectrum of SM is halved compared with the previous SM, which guarantees all possible combinations of input signals can be mapped to different wavelengths. By this means, the output wavelengths of $SM_2$ encompass four logic minterms of Signal A and Signal B: AB ($\lambda_1$, $\lambda_2$), A$\overline{B}$ ($\lambda_3$, $\lambda_4$), $\overline{A}$B ($\lambda_5$, $\lambda_6$), $\overline{A}\overline{B}$ ($\lambda_7$, $\lambda_8$). And after $SM_3$, eight logic minterms of three input operands (A, B, C) are carried on the light at eight different wavelengths. According to this rule, the $2^N$ logic minterms of $N$ inputs can be generated with $2^N$ wavelength channels and only $N$ input modulators are required. After generating all logic minterms, a WS is placed to assemble the wavelength channels. In accordance with the minterm components of targeted logic function, the accurate logic operation will be attained by selecting corresponding wavelength channels.

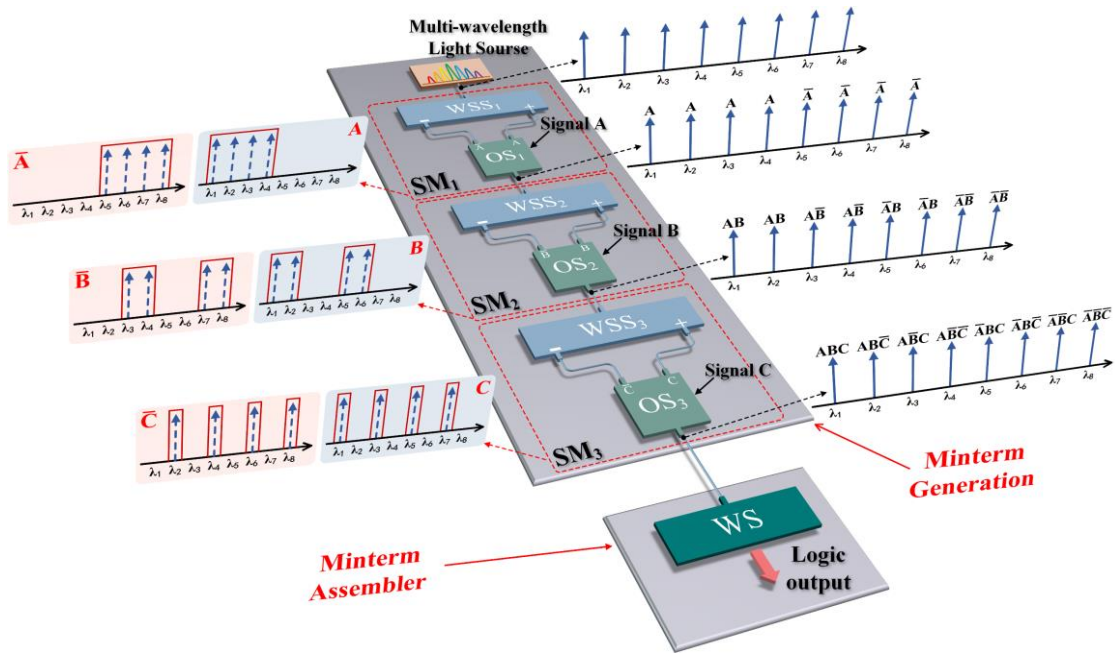

Fig.1 Operating principle of the proposed PLA based on parallel spectrum modulation. SM, spectral modulators; WSS, wavelength selective switch; OS, optical switch; WS, waveshaper.

### Experiment

Given that the WSS in the experiment allows for wavelength tuning within approximately 40nm around C-band, we opt for 256 wavelength channels with an interval of 0.15 nm to achieve eight-input PLA, which consists of 8 cascading SMs for minterm generation and a WS for minterm combination. The details of experimental setup are shown in Supplementary 1. Fig. 2(a) depicts the schematic diagram for the proposed PLA. The minterm generation part consists of eight cascading SMs, which needs 256 eight-input AND gates to generate complete logic minterms. Here, the method of parallel spectrum modulation fully leverages the abundant wavelength resources of light, leading to a substantial reduction in the number of devices. The generated minterms are at different wavelength channels represented by the blue lines in Fig. 2(a). All minterms are generated in one optical path, which also simplifies the spatial complexity. As for the configuration part, whether to use the minterms depends on the selection of corresponding wavelength channels.

We firstly demonstrate the capacity of minterm generation (8-256 decoder) of the PLA. The measured confusion matrix in Fig. 2(b) illustrates that each state of eight input operands corresponds to a wavelength channel ($I_8...I_1$ has 256 states ranging from 0 to 255 in the decimal form). It also indicates each input wavelength channel is independent and represents one logic minterm. And the extinction ratio between each wavelength channel exceeds 17dB. Owing to the background noise of erbium doped fiber amplifier (EDFA) placed between the fourth and fifth SM, some wavelength channels will appear noise spikes shown in Fig. 2(c). In the short wavelength region, the efficiency of EDFA is comparatively diminished and the background noise increases, resulting in relatively lower extinction ratio for theses wavelength channels. The 256 wavelength channels have $2^{256}$ possible combinations, indicating the proposed PLA possesses $2^{256}$ diverse logic functions. And it can guide optical designs for a wide variety of logic operations. Fig. 2(d) depicts four experimental examples for arbitrary logical output, where four letters of "HUST" are generated in the two-dimensional matrices composited by $I_8I_7I_6I_5$ and $I_4I_3I_2I_1$, demonstrating the programmable capability of PLA.

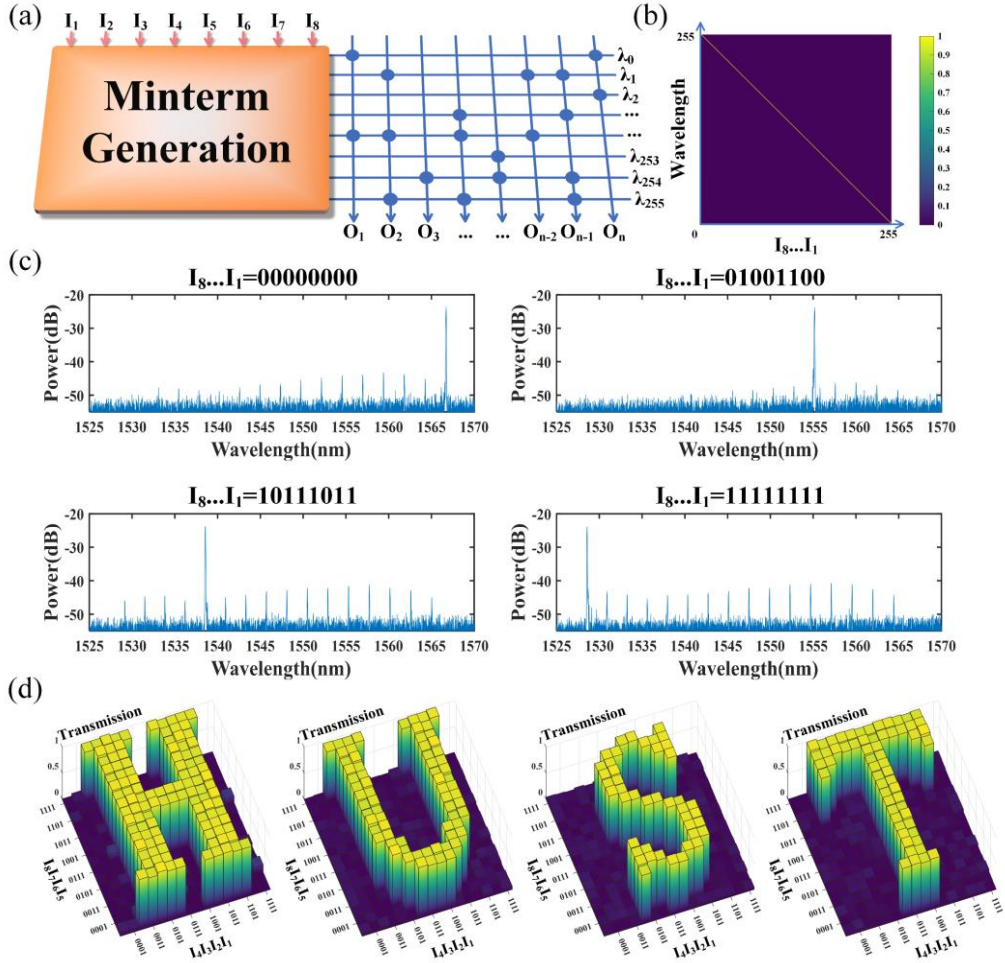

Fig.2. Reconfigurable capacity of PLA. (a) Schematic diagram for the proposed PLA; (b) Measured confusion matrix between 256 input states and output power of 256 wavelength channels. (c) Output spectrums of PLA for four different minterms. (d) Experimental examples for arbitrary logical output, where four letters of "HUST" are generated in the two-dimensional matrices composited by $I_8I_7I_6I_5$ and $I_4I_3I_2I_1$.

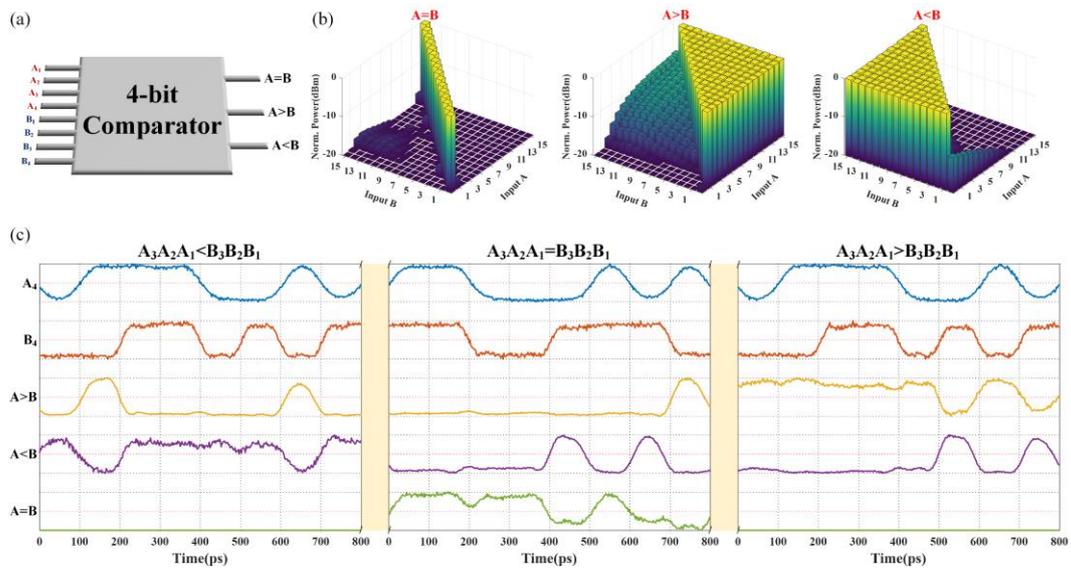

Fig.3 Experimental results of the 4-bit comparator. (a) The schematic diagram of 4-bit comparator. (b) Power distributions of three output ports between input Signal A ($A_4A_3A_2A_1$) and Signal B ($B_4B_3B_2B_1$). (c) Input and output waveforms of the comparator in the situations of $A_3A_2A_1 < B_3B_2B_1$, $A_3A_2A_1 = B_3B_2B_1$, $A_3A_2A_1 > B_3B_2B_1$.

Apart from the logic functions mentioned above in previous experiments, some common functional logic devices are further proved based on the PLA. We configure the PLA to implement a 4-bit comparator schematized in Fig. 3(a). The input Signal A ($A_4A_3A_2A_1$) and Signal B ($B_4B_3B_2B_1$) can both take a total of 16 integers ranging from 0 to 15 in the decimal form. Fig. 3(b) depicts the results of comparison between Signal A and Signal B, whose extinction ratio between logic high and low levels is more than 9dB. The extinction ratio depends on the quantity and quality of logic minterms contained in the output. Since function of A<B contains more <span style="color:red">minterms in the long wavelength channels,</span> the performance is slightly worse than function of A>B. After validating the functions of comparator in all input states, the high-speed optical switches are utilized to load $A_4$ and $B_4$ at 10Gb/s. Fig. 3(c) presents the waveforms of input Signal $A_4$, Signal $B_4$, and the output results in different situations between $A_3A_2A_1$ and $B_3B_2B_1$. The accurate execution of corresponding logic function at 10Gb/s demonstrates the high-speed computing capability of the proposed PLA (More details are in Supplementary 1).

Other functions like 4-bit adder and multiplier schematized in Figs. 4(a) and 4(b) are also realized by the PLA. The extinction ratio between high and low levels is over 10 dB (Power distributions of output ports are presented in Supplementary 2). A decision threshold is set to the binary data from experimental results. The output values of 4-bit adder and multiplier in Figs. 4(c) is derived by converting the binary data into decimal format. As long as each out port executes right logic operations, the final results will be accurate, which is assured thanks to satisfactory extinction ratio.

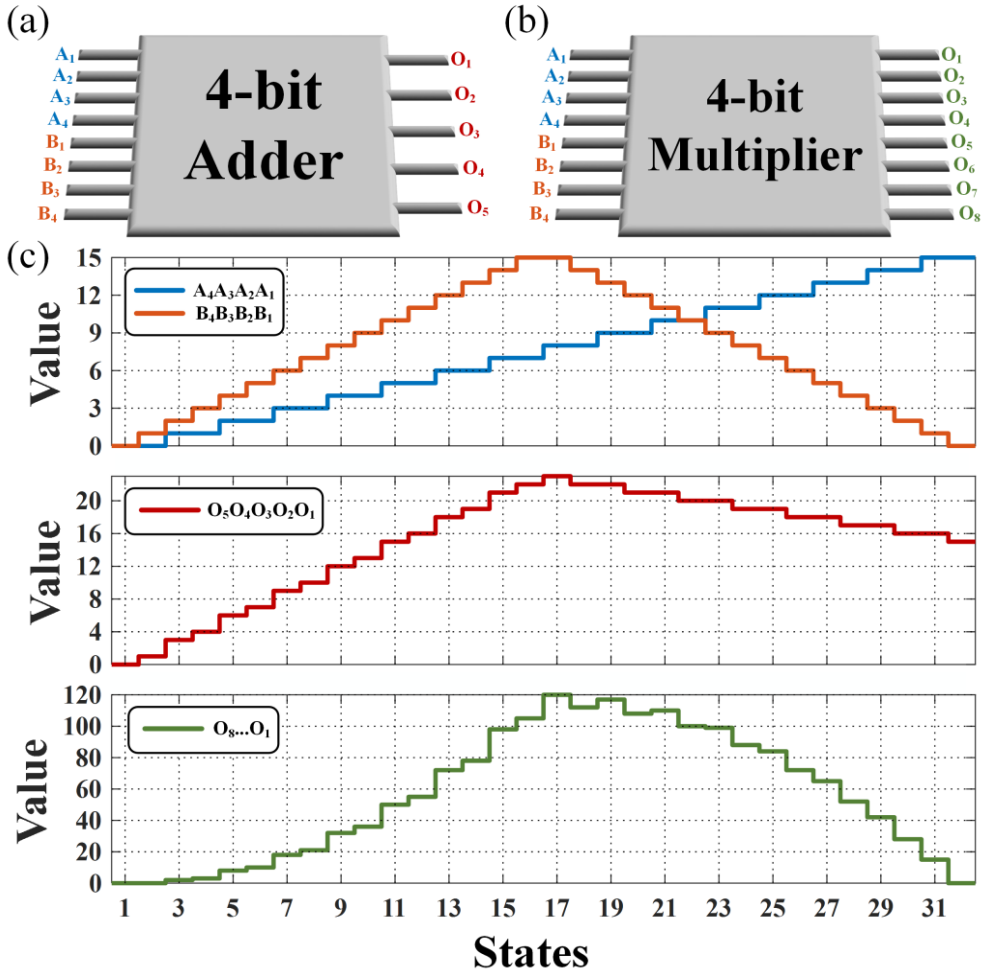

Fig.4 Experimental results of 4-bit adder and multiplier. The schematic diagrams of (a) 4-bit adder and (b) 4-bit multiplier. (c) The decimal values of Signal A ($A_4A_3A_2A_1$) and Signal B ($B_4B_3B_2B_1$), output results of 4-bit adder ($O_5O_4O_3O_2O_1$) and 4-bit multiplier ($O_8...O_1$) in decimal form.

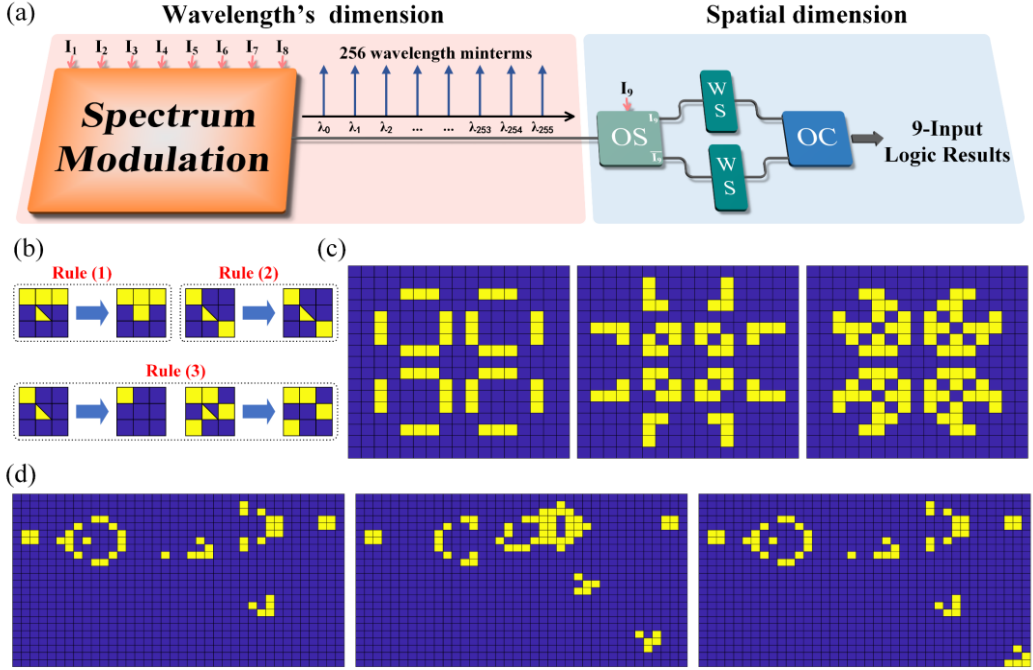

Fig.5 Optical platform for two-dimensional cellular automaton. (a) Nine-input' PLA realized by combining the wavelength's and spatial dimensions. (b) The evolutionary rules of Conway's Game of Life[34], in which the two different colors represent the cell's state (yellow represents the live state and blue represents the dead state). (c) Pulsar-explosion-like patterns and (d) Glider-gun-like patterns of cell's evolution in Conway's Game of Life implemented by the 9-bit PLA. OC, optical coupler.

Constrained by the resolution and operating wavelength range of WSS in the experiment, 256 wavelengths are applied to realize eight-input PLA. Here we expand to the nine-input PLA by adding a 1×2 optical switch in the spatial dimension shown in Fig. 5(a). The 256 wavelength minterms are first generated by eight cascading SMs, then modulated by 1×2 OS. As a result, the upper port output of OS contains 256 wavelength channels representing minterms with Signal $I_9$, and the lower port outputs 256 minterms with $\overline{I_9}$, which includes all the minterms generated by nine operands. Two WSs are followed behind the OS to select the required minterms, respectively. And the selected wavelength channels are mixed together by an optical coupler, thereby realizing nine-input logic functions (More details in Supplementary 3). The ability of large-scale input of PLA shows great potential for complex phenomenon simulation, such as CA.

Employing interactions with surrounding cells, CA can manifest various complex behaviors[35], which finds widespread applications in numerous fields like simulating the traffic flow[36,37], forest fire[38], the physical and chemical phenomena[39,40], etc. However, the existing optical platform can only support the simulation of one-dimensional CA and requires electrical threshold judgement[41]. Based on the nine-input PLA, we realize the optical two-dimensional CA for the first time and perform the Conway's Game of Life, whose optical output can directly represent the state of the cell (live or dead) in the next iteration. As one of the most classical demonstrations by CA, the Conway's Game of Life are designed to simulate the cells' evolutionary process[34]. The rules are shown in Fig. 5(b), given by

*(1) if the center cell is surrounded by three live cells, it becomes live in the next iteration, regardless of its current state;*

*(2) if there are two live cells around the center cell, it stays in its present state in the next iteration;*

*(3) if the center cell is surrounded by less than two or more than three live cells, it becomes dead in the next iteration, regardless of its current state.*

We encode these rules into the PLA and utilize it to predict the state of each cell in the next iteration. Figs. 5(c) and (d) shows two different kinds of the cells' evolution (pulsar explosion and glider gun), which is consistent with the results simulated on the computer (truth table for the proposed CA is shown in Supplementary 4). Compared with the $256(2^{2^3})$ possible modes of one-dimensional CA[42], the two-dimensional CA can support $2^{512}$ ($2^{2^9}$) possible modes, indicating the great enhancement of optical CA's functionalities.

## Discussion

Table 1 shows the comparison between the proposed eight-input PLA and existing methods, including AO and EO PLAs. The number of modulators is linearly related to the number of input operands in the proposed PLA, which makes it practical to support more inputs. The notable expansion of generated minterms and functions highlights the tremendous advancement of optical PLA. Moreover, the eight-input PLA isn't its ultimate potential. The scale of the proposed PLA can be further extended by broadening the range of wavelength or narrowing the interval between wavelengths to apply more channels. Note that the decrease of wavelength's interval will lead to the reduction of operating speed owing to the adequate bandwidth requirements for high-speed signal. Consequently, there is a trade-off between speed and the number of operands. Here we give the estimation of PLA's scale when the wavelength range is 1500-1600nm and operating speed is 1Gbit/s. The total bandwidth can be written as,

$$\Delta f = \frac{c}{\lambda_1} - \frac{c}{\lambda_2} = 12491.3\text{GHz} ,$$

where $\lambda_1$=1500nm and $\lambda_2$=1600nm. Therefore, the maximum number of wavelength channels (*W*) it can support is about 12491. The number of input operands *N* is given by

$$N = [\log_2 W] = 13 ,$$

which is a considerable scale for programmable logic devices. In addition, the scale of PLA can be further extended by employing more broadband light or making full use of the dimensions of wavelength and space.

Table 1: Comparison between the proposed PLA and existing methods

| Method | Input operands | Minterms | Functions | Modulators for N operands |
|---|---|---|---|---|
| AO PLA[30] | 3 | 8 | 256 | -------------- |
| EO PLA[32,33] | 4 | 16 | 65536 | N²(Multi high levels) |
| Proposed PLA | 8 | 256 | $2^{256}(\sim10^{77})$ | N |

It's also optimistic to achieve the integration of SM on silicon platform. As for the periodic transmission spectrum, the ring-assisted Mach Zehnder interferometer (MZI) can be a favorable selection[43]. And the MZI structure can be exploited to perform high speed 1×2 optical switch with an wide bandwidth[44]. These two integrated structures can construct the SM and realize the generation of logic minterms.

## Conclusion

In conclusion, we implement an eight-input optical PLA based on parallel spectrum modulation. The generated 256 minterms have $2^{256}$ possible combinations, revealing its extensive range of functionalities. The 8-256 decoder, 4-bit comparator, adder and multiplier are experimentally demonstrated by configuring the proposed PLA. And we verify its high-speed computing capability at 10Gb/s. Moreover, we extent to nine-input PLA via the introduction of spatial dimension and realize the optical two-dimensional CA for the first time. The significant expansion of scale of optical PLA and new optical logic applications create a new avenue for the advancement of optical digital computing.

## Methods:

### Experimental setup

The concrete experimental setup is shown in Fig. S1. The broadband continuous light is first generated by the broadband light source and then shaped by a WS thus generating 256 beams in wavelength channels with 0.15nm interval. The SM is composed of a WSS and a 1×2 optical switch. By cascading eight SMs, the 256 wavelength channels can load different logic minterms. As for the high-speed experiment, the signals are generated by a bit pattern generator and the results are detected by a 50GHz photodetector. Owing to the power jitter in single wavelength channel generated by broadband light source, the input light in wavelength channels correspond to the logic minterms of targeted logic functions is replaced by the lasers' outputs.

### Acknowledgements:

This work was partially supported by the National Key Research and Development Project of China (2022YFB2804201), the National Natural Science Foundation of China (U21A20511, 62075075, 62275088), the Innovation Project of Optics Valley Laboratory (Grant No. OVL2021BG001).

### Authors' contributions

WZ, WB and HZ conceived the idea. WZ, BW, WG designed and performed the experiments. HZ, JD, WZ and BW discussed and analyzed data. WZ prepared the manuscript. HZ, JD, DH and PW revised the paper and XZ supervised the project. All authors contributed to the writing of the manuscript.

### Data availability

All the data related to this paper are available from the corresponding authors upon request.

**Additional information:** There is no additional information for this article.

**Competing financial interests:** The authors declare no competing financial interests.

# Supplementary materials for "Eight-input optical programmable logic array enabled by parallel spectrum modulation"

Wenkai Zhang[1†], Bo Wu[1†], Junwei Cheng[1], Hailong Zhou[1*], Jianji Dong[1,*], Dongmei Huang[2,3], P. K. A. Wai[4] and Xinliang Zhang[1]

[1]Wuhan National Laboratory for Optoelectronics, School of Optical and Electronic Information, Huazhong University of Science and Technology, Wuhan 430074, China

[2]The Hong Kong Polytechnic University Shenzhen Research Institute, Shenzhen 518057, China

[3]Photonics Research Institute, Department of Electrical Engineering, The Hong Kong Polytechnic University, Hong Kong, 999077, China

[4]Department of Physics, Hong Kong Baptist University, Kowloon Tong, Hong Kong, 999077, China

*Corresponding author: hailongzhou@hust.edu.cn; jjdong@hust.edu.cn

[†]These authors contributed equally to this work

## S1: Experimental setup of the optical programmable logic array

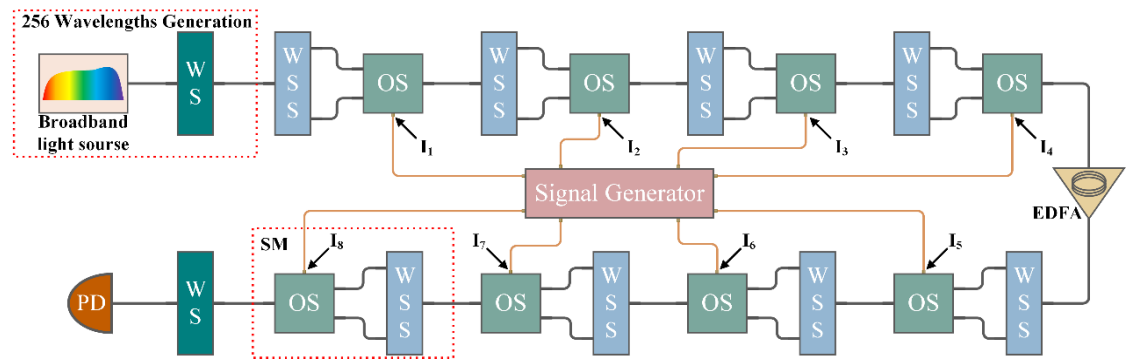

Fig. S1 Experimental setup of the eight-input programmable logic array (PLA) based on parallel spectrum modulation. WS, waveshaper; WSS, wavelength selective switch; OS, optical switch; SM, spectral modulators; EDFA, erbium doped fiber amplifier; PD, photodetector.

The experimental implementation of the eight-input PLA is presented in Fig. S1. Here, we use a broadband light source and a WS to generate 256 beams with 0.15nm interval. The 256 beams are then modulated by eight cascaded SMs which is composed of a WSS and a 1×2 OS. Owing to the power loss induced by optical components, we add an EDFA between the fourth and fifth SMs. Following the SMs, a WS is used to select targeted wavelength channels and the final logic computing results are detected by a PD.

To validate the high-speed computing capacity of the proposed PLA, we load high-speed signals at 10Gbit/s to the OSs in fourth and eights SMs (Signals $A_4$ and $B_4$ of the four-bit comparator). Owing to the jitter of single wavelength generated by broadband light source, the beams corresponding to the logic minterms of targeted logic functions are replaced by the lasers' outputs.

## S2: Power distributions of four-bit adder and multiplier

Figs. S2 and S3 show the power distributions of four-bit adder and multiplier, respectively. The extinction ratio between high and low levels exceeds 9dB. The distinct contrast of these two levels guarantees the accuracy of the final computation results.

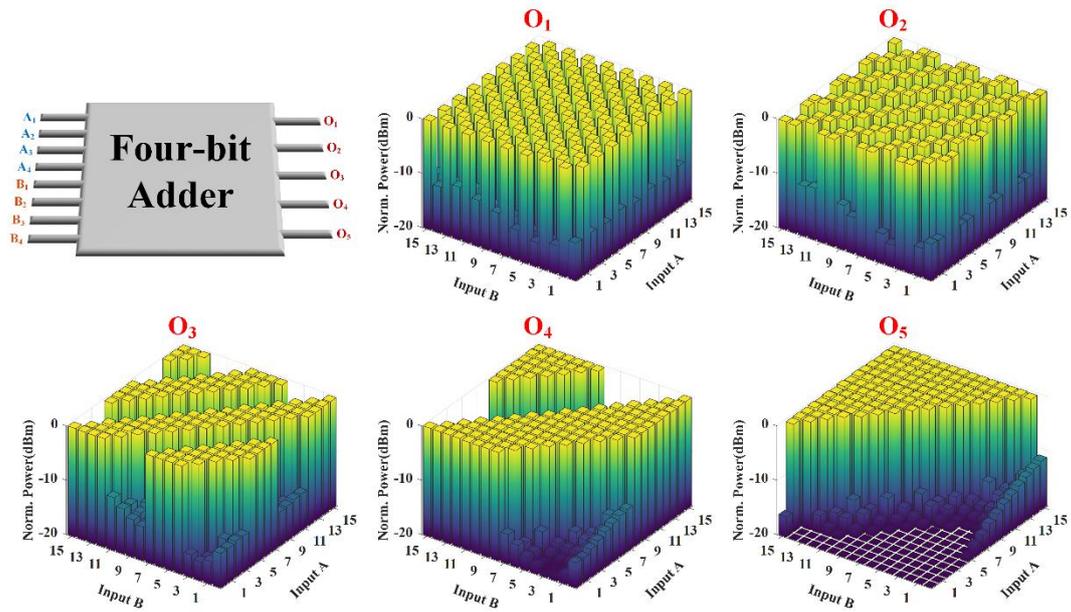

Fig. S2 Four-bit adder's power distributions of five output ports ($O_5O_4O_3O_2O_1$).

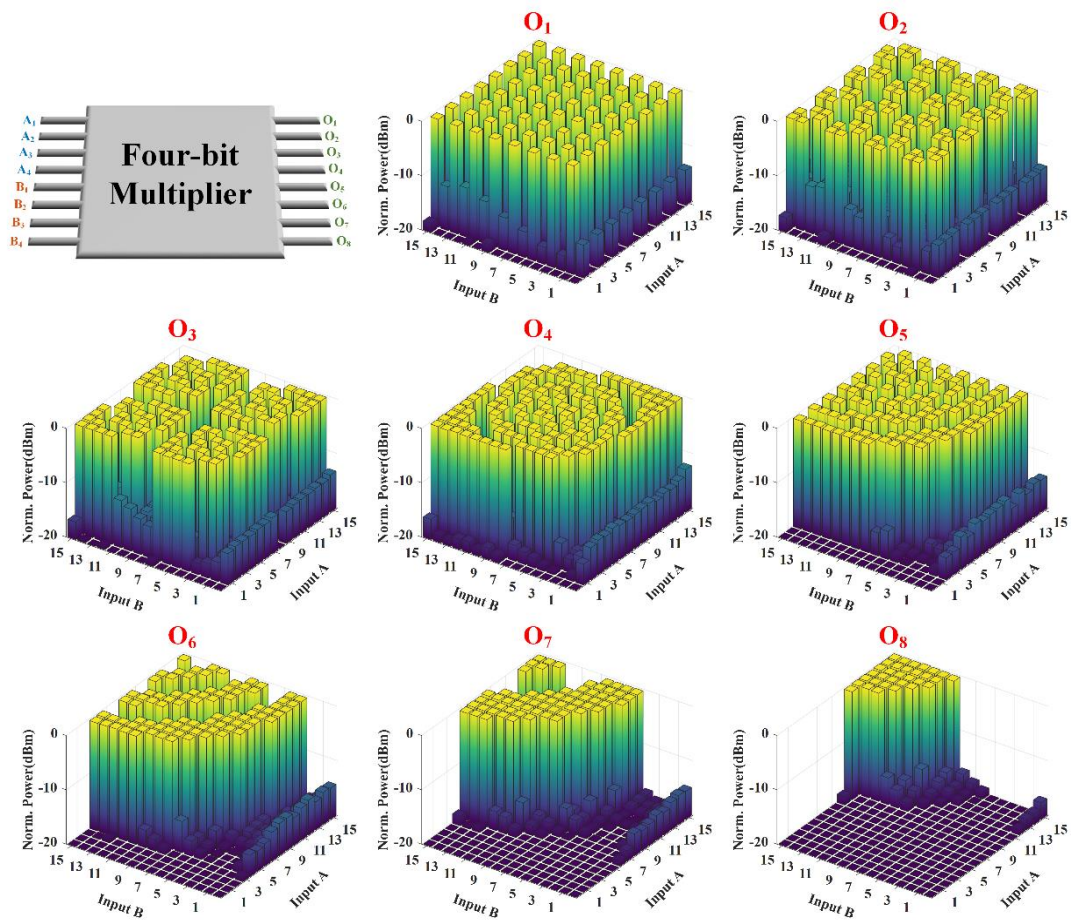

Fig. S3 Four-bit multiplier's power distributions of eight output ports ($O_8...O_1$).

**S3: PLA's expansion principle of the combination between wavelength's and spatial**

**dimensions**

In this section, we give a feasible way to expand the PLA's scale by combining wavelength's and spatial dimensions. For simplicity, the schematic diagram of four-input PLA is shown in Fig. S4(a). In the wavelength's dimension, we load the first two operands (A, B) to the two cascaded SMs and obtain four minterms ($M$(A, B)) represented by four wavelengths. The $M$(A, B) is then input into a 1×2 OS, whose upper channel outputs C$M$(A, B) and lower channel outputs $\overline{C}M$(A, B). And the two output ports are followed by two OSs to load the operand D. As a result, the four ports will output 16 minterms of 4 input operands. To explain more clearly, Fig. S4(b) present the correspondence between generated minterms and the two used dimensions. The 16 minterms correspond to the 16 possible combinations between four wavelengths and four spatial ports. We can control the four WSs after the four output ports to select minterms and realize targeted logic operations.

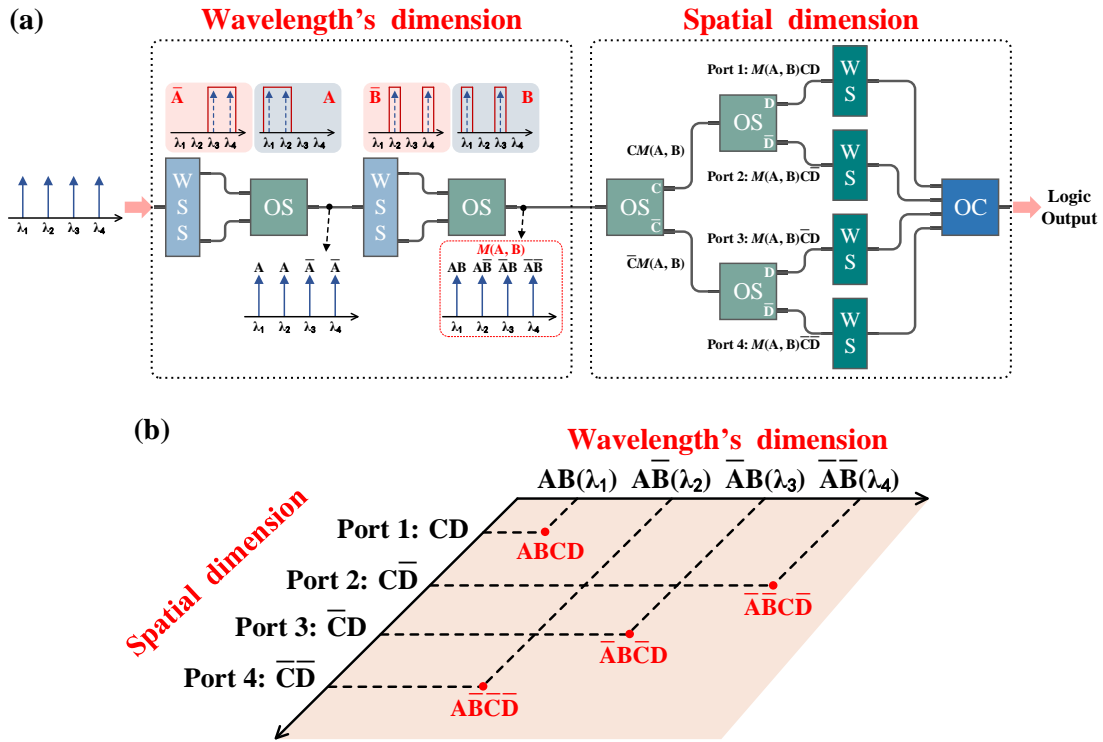

Fig. S4 PLA's expansion principle of the combination of wavelength's and spatial dimensions. (a) Schematic diagram of four-input PLA by combining the wavelength's and spatial dimensions. (b) The correspondence between generated minterms and the two dimensions.

**S4: The measured truth table for two-dimensional cellular automaton**

Fig. S5 demonstrates the power distribution (measured truth table) for two-dimensional cellular automaton based on the nine-input PLA. It can be divided into two parts according to the operand $I_5$ (the state of center cell). If the center cell is dead in the current iteration ($I_5$=0), only when three surrounding cells are live (three of other eight operands' values are 1) the center cell can be live in the next iteration. And if the center cell is dead in the current iteration ($I_5$=1), it can maintain its live state when two or three surrounding cells are live (two or three of other eight operands' values are 1). The extinction ratio between the live and dead states of cell is over 9dB, which ensures the accuracy of cells' evolution in the experiment.

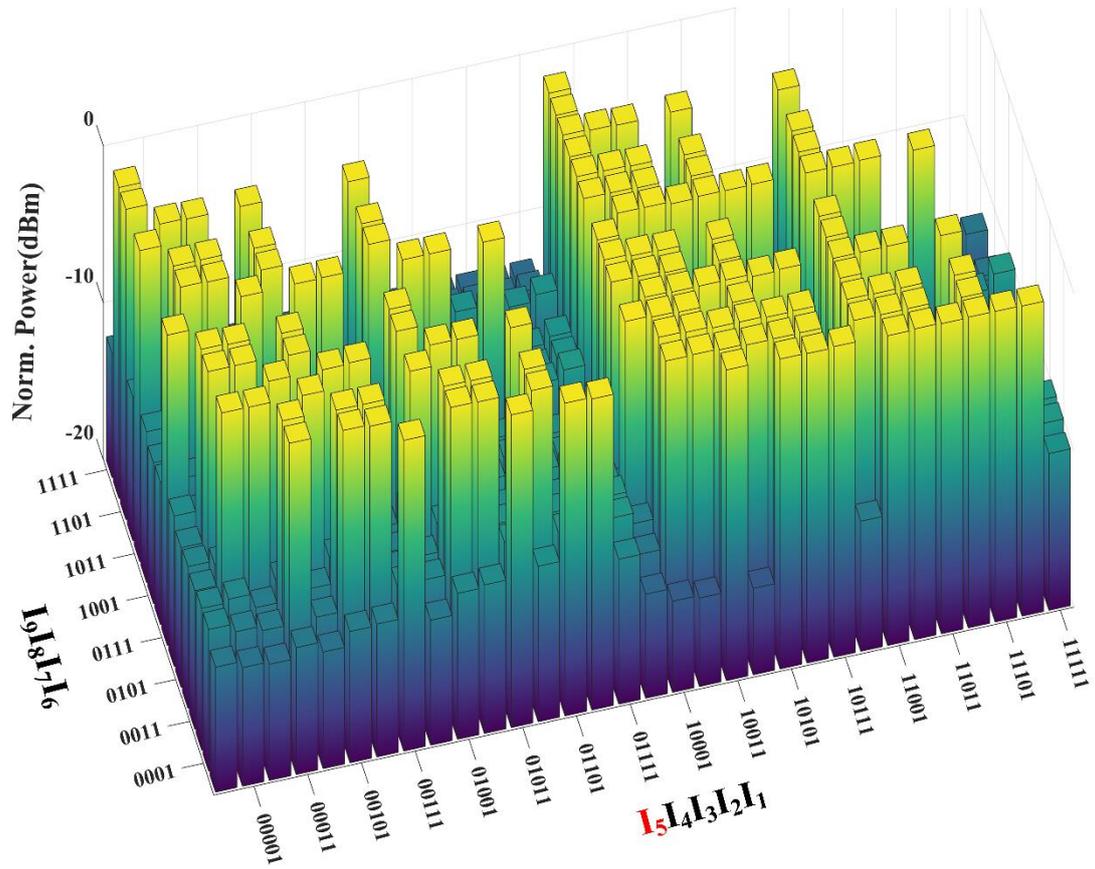

Fig. S5 Power distribution (measured truth table) for two-dimensional cellular automaton. Operand $I_5$ corresponds to the state of center cell, and other operands correspond to the surrounding cells' states.